\documentclass[namedreferences,hyperref,optionalrh]{spr-sola}
\usepackage{graphicx}        
\usepackage{color}           
\usepackage{ulem}            

\newcommand\new[1]{#1}
\newcommand\old[1]{}

\chardef\us=`\_

\newcommand{\arcsec}{$^{\prime\prime}$}
\newcommand{\arcmin}{$^\prime$}
\newcommand{\degree}{$^\circ$}

\begin{document}

\begin{frontmatter}
\title{CATEcor: an Open Science, Shaded-Truss, Externally-Occulted Coronagraph}

\author[addressref={aff1},corref,email={craig.deforest@swri.org}]{\inits{C.E.}\fnm{Craig}~\snm{DeForest}\orcid{0000-0002-7164-2786}}
\author[addressref=aff1]{\inits{D.B.}\fnm{Daniel B.}~\snm{Seaton}\orcid{0000-0002-0494-2025}}
\author[addressref=aff1]{\inits{A.}\fnm{Amir}~\snm{Caspi}\orcid{0000-0001-8702-8273}}
\author[addressref=aff1]
{\inits{M.}\fnm{Matt}~\snm{Beasley}}
\author[addressref=aff4]{\inits{S.}\fnm{Sarah~J.}~\snm{Davis}\orcid{0009-0008-4901-0601}}
\author[addressref=aff1]
{\inits{N.F.}\fnm{Nicholas~F.}~\snm{Erickson}\orcid{0000-0001-6028-1703}}
\author[addressref=aff1]
{\inits{S.A.}\fnm{Sarah~A.}~\snm{Kovac}\orcid{0000-0003-1714-5970}}
\author[addressref=aff1]
{\inits{R.}\fnm{Ritesh}~\snm{Patel}\orcid{0000-0001-8504-2725}}
\author[addressref=aff1]
{\inits{A.}\fnm{Anna}~\snm{Tosolini}\orcid{0009-0000-3510-8116}}
\author[addressref=aff1]
{\inits{M.J.}\fnm{Matthew J.}~\snm{West}\orcid{0000-0002-0631-2393}}
\runningauthor{DeForest et al.}
\runningtitle{CATEcor: a Shaded-Truss External Coronagraph}

\address[id=aff1]{Southwest Research Institute, Boulder, Colorado, USA}
\address[id=aff4]{Independent Researcher}

\begin{abstract}
We present the design of a portable coronagraph, CATEcor, that incorporates a novel
``shaded truss'' style of external occultation and serves as a proof-of-concept for that family of coronagraphs.  The shaded truss design style has 
the potential for broad application in various scientific settings. We conceived CATEcor 
itself as 
a simple instrument to observe the corona during the darker skies available during
a partial solar eclipse, or for 
students or interested amateurs to detect the corona under 
ideal non-eclipsed conditions.  CATEcor is therefore optimized for simplicity and
accessibility to
the public.  It is implemented using 
an existing dioptric telescope and an adapter rig that mounts in front of the 
objective lens, restricting the telescope aperture and providing external 
occultation. The adapter rig, including occulter, is fabricated using fusion
deposition modeling (FDM; colloquially ``3D printing''), greatly reducing cost.  The 
structure is designed to be integrated with moderate care and may be replicated in 
a university or amateur setting.  While CATEcor is a simple demonstration unit,
the design concept, process, and trades are useful for other more sophisticated 
coronagraphs in the same general family, which might operate under normal daytime skies outside the annular-eclipse conditions used for CATEcor.
\end{abstract}
\begin{motto}
\textit{
Sometimes, the truth is found not in seeking, but in hiding. -- Khaled Hosseini
}
\end{motto}

\end{frontmatter}


\section{Introduction} \label{sec:intro}

\begin{figure}[tb]
    \centering
    \includegraphics[width=3.5in]{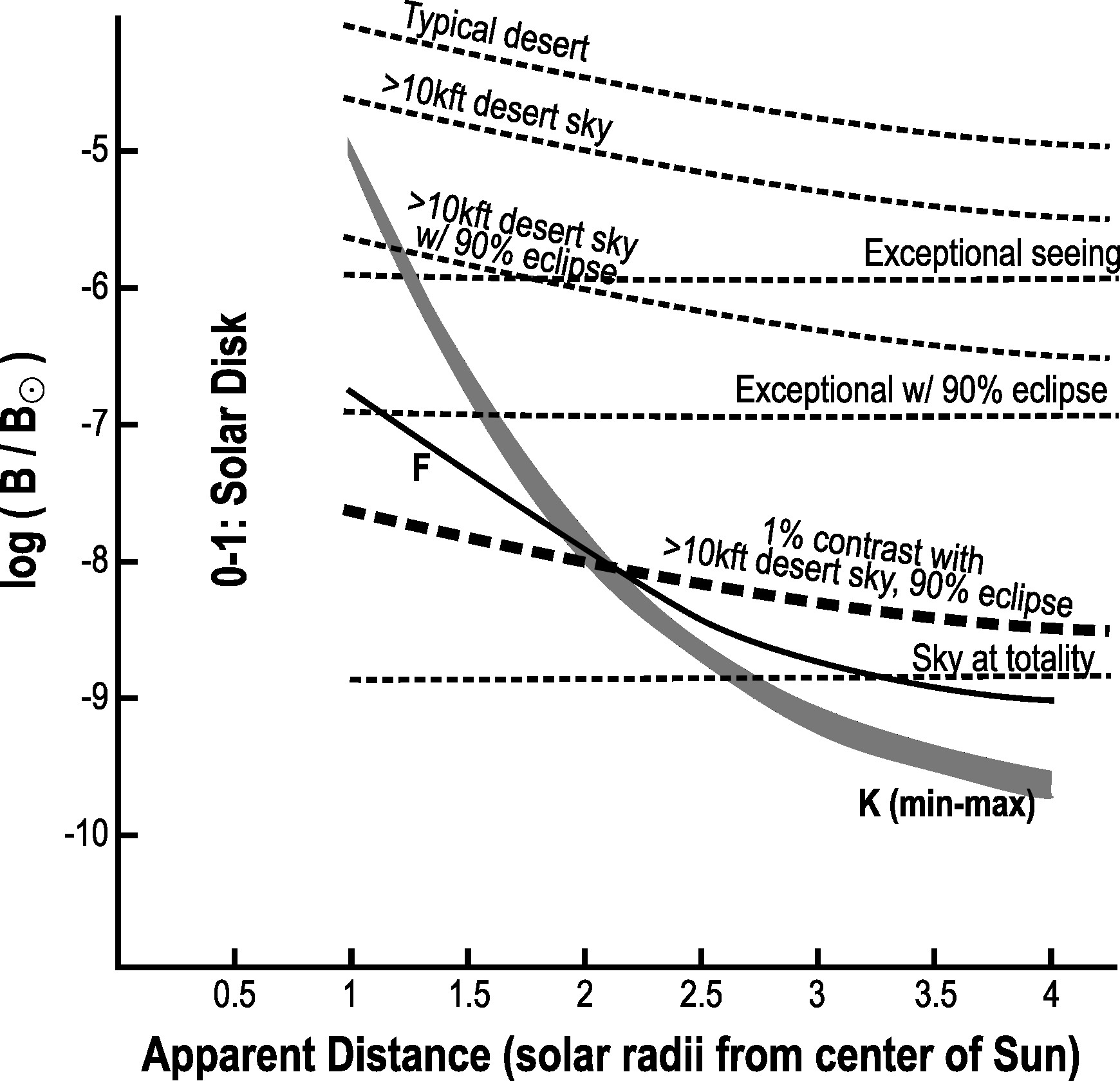}
    \caption{Ground-based solar coronagraphs must contend with not only instrumental scatter but the background sky
    brightness (dotted lines for various conditions), which falls off slowly compared with the brightness of the K-corona itself (grey line).  This drives coronagraph solutions that are optimized for the very low corona, 
    requiring internal occultation.  In very high-altitude conditions or during an annular solar eclipse, the sky 
    brightness is reduced by an order of magnitude or more, enabling externally occulted coronagraphs to image the 
    corona to 3 or more solar radii.  Figure adapted with permission from \citet{golub_pasachoff_2009}.}
    \label{fig:sky-brightness}
\end{figure}

Solar coronagraphs work by blocking out the Sun to produce an ``artificial eclipse'' \citep{lyot_1930}, allowing imaging
of the corona itself around the bright Sun.  Ground-based instruments must contend not only with instrumental scattering 
but also with sky brightness (Figure~\ref{fig:sky-brightness}). Such instruments are optimized to image the innermost 
portion of the corona, which rises above the local sky brightness; they include the Mauna Loa K-coronameter \citep{Altschuler_Perry_1972,Fisher_1981} and more modern instruments including CoMP \citep{tomczyk_2009} 
and K-Cor \citep{dewijn_2012}.  Instruments that image low coronal altitudes generally use internal occultation, in which an initial focusing 
optic produces a real image of the Sun, and the light in that image is rejected from the instrument by a physical object or a hole 
in a reflecting optic.  Internal occultation allows very precise selection of which altitudes within the corona will be imaged, 
at the cost of requiring a very-low-scatter initial optic and generating instrumental stray light in the far field.

Spaceborne coronagraphs have practically no limitation from sky brightness, and can image the corona much farther from 
the
Sun,
by using 
external occultation to reject the bulk of the Sun's rays even before the first optic.  This greatly reduces stray light in
the far field, at the cost of vignetting the field near the Sun itself. Examples of externally occulted coronagraphs include 
LASCO/C3 \citep{brueckner_etal_1995} and STEREO/COR2 \citep{howard_etal_2008}.  The most recent
generation of spaceborne coronagraphs, including CCOR \citep{thernisien_2021} and PUNCH/NFI \citep{colaninno_2019}, eschew
internal or secondary occultation altogether and rely entirely on a highly engineered external occulter to reduce 
stray light to levels compatible with coronal imaging as far as 30 apparent solar radii ($R_\odot$) from the Sun.

The limitations of ground-based coronagraphs are greatly mitigated during conditions that
significantly improve (reduce)
sky brightness.  Figure~\ref{fig:sky-brightness} shows typical sky brightness curves under 
various conditions.  During an annular solar eclipse, or at exceptionally high altitude in the 
atmosphere (e.g., above 80,000 ft. altitude), sky brightness can be reduced by an order 
of magnitude or more compared to high desert conditions; in turn, that may enable 
imaging of the middle corona \citep{west_2023} at or above 3~R$_\odot$ from the
Sun without flying an instrument into space or requiring a total solar eclipse.  This insight
led us to consider new instrument designs that might enable ground-based imaging of the 
middle corona, farther from the Sun than is possible from the ground under normal conditions.

The imaging quality required to capture the middle corona is modest by the standards of either 
spaceborne coronagraphs (which are optimized for very low stray light and small instrument 
dimensions) or ground-based coronagraphs (which are
optimized to image very close to the Sun).  Hence we did not consider a complete new instrument, but
a module that could extend an existing, well-defined imaging system.  We adopted the telescope and camera from the Citizen Continental-America Telescopic Eclipse 2024 (CATE24) program \citep{Caspi_etal_2023,Patel_etal_2023}, which in turn is a next-generation version of the original Citizen CATE \citep{CATE_paper_2020} project carried out in 2017.  The CATE24 observing equipment comprises a Long Perng doublet refracting telescope with 80~mm aperture and 500~mm focal length (f/6.25) on a German Equatorial Mount, and a FLIR Blackfly polarizing CMOS camera with an integrated multiplex polarizer mask, allowing recovery of Stokes I, Q, and U across the entire field of view (FOV) from each exposure.  The imaging resolution is 2.9{\arcsec} FWHM, limited by diffraction from the aperture, and is matched to the 1.43{\arcsec}/pixel plate scale of the overall system. \citep[See][for further details.]{Patel_etal_2023}

CATEcor modifies the CATE24 telescope with a 
clip-on assembly 
comprising an aperture stop just in front of the telescope objective, 
and an external occulter 75~cm in front of the stop. The occulter is 
supported by a
truss that remains fully shaded by the occulter itself; this reduces 
scatter from the support structure,
and reduces or eliminates the need for a dark ``vestibule'' to support a pylon and
the occulter, as in existing externally
occulted instruments \citep[e.g.,][]{howard_etal_2008}.    CATEcor thus
embodies a new type of instrument, a ``shaded-truss externally-occulted 
coronagraph\footnote{The shaded-truss design is the subject of a patent 
application by Southwest Research Institute. First public disclosure was 
2023 October 6. Contact the authors for more information.}''.  Compared 
to conventional internally-occulted coronagraph designs, the 
shaded-truss approach has lighter weight, lower instrumental stray light in the far field, and the beneficial smooth inner-field sensitivity rolloff that is characteristic of other externally-occulted designs, at the cost of a more complex vignetting function (as the 
instrument looks \textit{through} a complex, structured truss).

The entire
CATEcor assembly may be 
integrated from readily obtainable parts and 3D printed elements, permitting individuals to reproduce CATEcor for hobbyist or student applications; thus we refer to CATEcor
as an ``open science'' instrument. 

In the following sections, we describe the trade space and requirements that
drive CATEcor, and present the actual design in sufficient detail to reproduce the 
instrument. 
Section~\ref{sec:requirements} gives requirements for CATEcor; Section~\ref{sec:design} presents the
design concept, develops specifications for the optomechanical elements, and presents the design 
elements.  Section~\ref{sec:integration} describes fabrication and integration steps for the instrument.  Section~\ref{sec:initial-testing} includes results from a full-Sun test observation. 
Section~\ref{sec:discussion} contains 
discussion of the novel design space of CATEcor and its relevance to future instrumentation, and 
we draw conclusions in Section~\ref{sec:conclusions}.  We also used CATEcor to image the corona at an 
annular eclipse; that observation is detailed by \citet{Seaton_etal_2024}.

\section{Requirements for CATEcor}\label{sec:requirements}

Two objectives drove us to build CATEcor: (1) to demonstrate coronal imaging with at-hand materials 
including a low-cost 
commercial telescope available to amateurs \citep[the CATE24 telescope;][]{Patel_etal_2023} and supplies 
and manufacturing technology available to 
hobbyists; and (2) to validate a novel, simplified coronagraph concept for coronal imaging.
The primary design requirements were:

\textbf{Overall structure}: CATEcor is intended as an easily manufacturable addition to the 
existing CATE24 telescopes, which may be directly mounted on the telescope and balanced with existing 
CATE24 project mounts, tripods, counterweights, etc. Alignment and calibration must be accessible to 
amateurs or students with the same skill level required of CATE24 operators in the main program.

\textbf{Imaging resolution}: The resolution requirement is driven by feature size in the corona.  Streamer tops are approximately 2--3{\arcmin} across, levying an azimuthal-direction resolution requirement of 2{\arcmin} at apparent distances of 1.8~R$_\odot$ or more from disk center.  This is modest or trivial for a small refractor such as the 
80~mm aperture CATE24 telescopes, but significant for an externally occulted coronagraph where the
effective aperture may be very small, and drives geometric aspects of the shaded-truss design.  A 2{\arcmin} diffraction limit requires an effective aperture roughly 1~mm across or more.  It also drives noise performance of the instrument, as in \citet{deforest_etal_2018}.

\textbf{Field of view}: CATEcor must be in principle able to image full circle from 1.5~R$_\odot$ 
to 2.5~R$_\odot$ under the dark sky associated with a 90\% annular eclipse, and must capture 
sufficient altitude range of the corona to unambiguously demonstrate imaging. The CATE24 telescopes have 
adequate FOV, extending to \new{at least} $3$~R$_\odot$\new{ in all directions}.  For an externally occulted design, the \old{limiting factor on the} inner edge 
of the \old{3D}\new{FOV} is \new{set by} the occulter/aperture geometry.  The \old{limiting factor on the} outer \old{edge}\new{limit}
of the effective FOV is \new{determined by} background brightness and associated noise characteristics. CATEcor
is expected to capture from close to 1.5~R$_\odot$ to roughly 2~R$_\odot$ with the possibility of a full
solar radius from 1.5--2.5~R$_\odot$, based on typical K-corona brightness (Figure~\ref{fig:sky-brightness}).

\section{The CATEcor design} \label{sec:design}

We conceived CATEcor as a front-end external occulter-and-aperture assembly (Figure~\ref{fig:ray-diagram}) to mount directly to the existing CATE24
telescopes, which are 80~mm diameter aperture tube-based dioptric telescopes
with 100~mm diameter retractable dust covers.  In keeping with the CATE24 project philosophy of reducing barriers
to scientific measurement, we specifically designed CATEcor to be easily assembled from 
readily available hardware and hobbyist Fusion Deposition Modeling (FDM) equipment (``3D printing'').
In principle, CATEcor could be duplicated by anyone with acccess to an amateur telescope, a computer, 
a 3D printer, and a hardware store.

\begin{figure}
    \centering
    \includegraphics[width=2.5in]{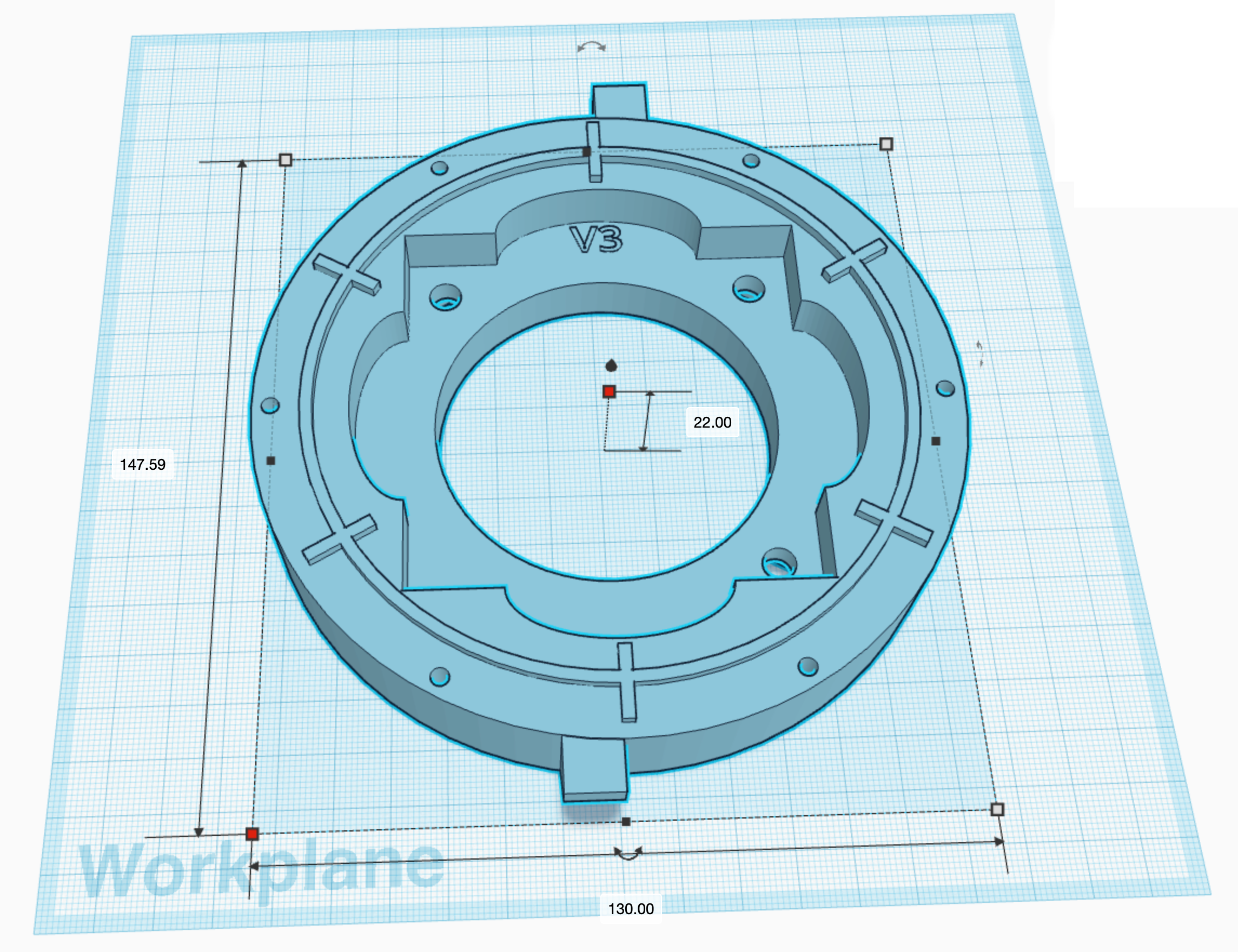}
    \caption{CATEcor front plate rests on top of the tube adapter.  A circular groove at the bottom mates with the circular ridge at the top of the tube adapter.  A central 55~mm diameter hole allows light to enter the telescope objective lens.  A square recess in the structure accepts an adjustable-iris aperture stop (not shown).  Radial ridges mate positively with radial recesses on the underside of the aperture piece, to ensure dimensional stability.  Hexagonal recesses (not visible) on the underside mate with M6 through-bolts to secure the adjustable-iris aperture stop. The top and bottom feature ``optical maze'' mounting rings both for alignment and to prevent stray light entering through joints in the assembly.  Small finger tabs on the perimeter help to mate/demate the press-fit alignment features with the aperture piece. }
    \label{fig:front-plate}
\end{figure}

\begin{figure}
    \centering
    \includegraphics[width=2.5in]{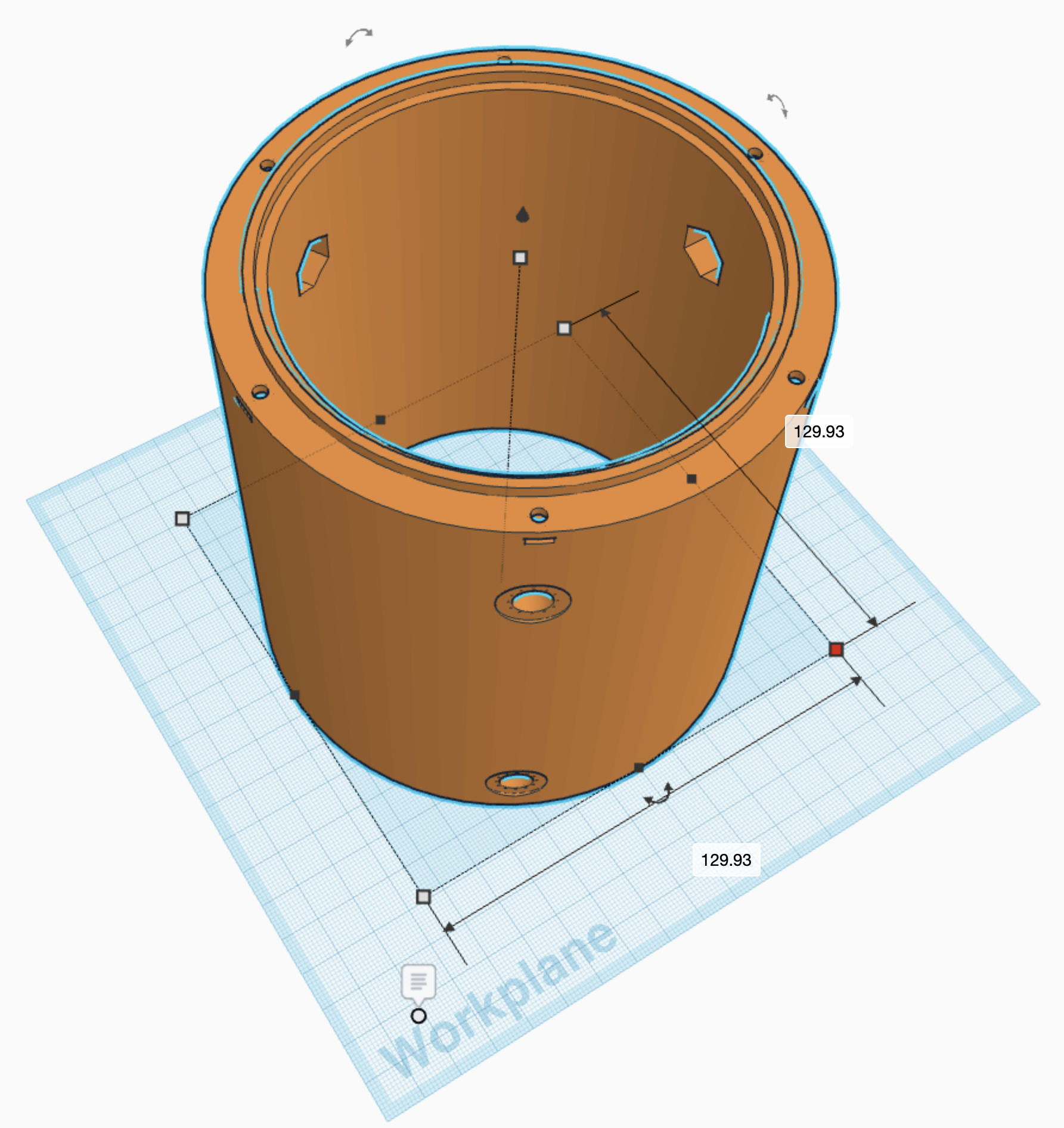}
    \caption{CATEcor tube extension is designed to slip over the dust cover of a CATE24 telescope.  It is secured by six nylon M8 bolts with captive hex nuts on the inside and nylon washers and hex nuts on the outside. A circular mount ring aligns the tube and prevents external light from entering.  The front plate and aperture are 
    secured by six M3 through-bolts into captive square nuts.}
    \label{fig:tube-adapter}
\end{figure}

The CATEcor assembly comprises fasteners and ancillary elements, linking four 3D printed parts: an external occulter (Figure~\ref{fig:occulter}),
an aperture piece (Figures~\ref{fig:aperture} and \ref{fig:aperture-drawing}), a front plate (Figure~\ref{fig:front-plate}), and a telescope tube extension (Figure~\ref{fig:tube-adapter}).  The occulter
is attached to the aperture piece by a 75~cm carbon-fiber hexapod truss made from commercially available
carbon-fiber rods, which are glued into 3D printed features in the occulter and aperture piece.

The telescope tube extension, \old{top}\new{front} plate, and aperture \old{front-end}\new{piece} were all designed using
the extremely simple web-based tool TinkerCAD \citep{TinkerCAD}.  TinkerCAD was ideal
because it is specifically designed to lower barriers to CAD design by
hobbyists and students, highlighting the accessibility of appropriate
mechanical design tools for everyone.  
The largest drawback of TinkerCAD for 
this application is that it it uses a coarse polyhedral tesselation for all curved surfaces.  
Typical outside dihedral angles produced by TinkerCAD, when approximating round surfaces, are 6{\degree} or more.  This was
acceptable for the mechanical interface structures such as the telescope attachment, but unacceptable
for the occulter
itself.  We therefore used the open-source FreeCAD software
\citep{FreeCAD} for the occulter itself.

All of the 3D printed parts' designs are freely available in ``.stl''
and other formats, and are available for download \citep{DeForest_2023}.

In the following subsections, we describe and develop the fundamental design concept of a 
shaded-truss externally occulted coronagraph (Section~\ref{sec:concept}); develop the 
occulter design (Section~\ref{sec:occulter}); and describe the designs for the truss 
(Section~\ref{sec:shaded-truss}), aperture (Section~\ref{sec:aperture}), front plate 
(Section~\ref{sec:front-plate}), and telescope extension tube (Section~\ref{sec:telescope-tube}).

\subsection{Shaded-truss externally occulted coronagraph concept design\label{sec:concept}}

The coronagraph design space is dominated by the observed strong radial gradient in the K-coronal 
brightness (Figure~\ref{fig:sky-brightness}).  Coronagraphs generally capture an 
annular FOV.  The inner edge is limited by the geometry and diffraction
characteristics
of the occulter.  The outer edge is limited by
background sources of light:  sky brightness for ground-based instruments or 
instrumental stray light for spaceborne instruments.  At large apparent distances
(elongations), the desired K-corona signal
drops well below other background sources including the F-corona; but
digital post-processing enables separation of the K signal even at very large 
elongation angles and contrast ratios as low as a few $\times~10^{-4}$
\citep{Jackson_1985,deforest_etal_2011}, enabling both wide-field coronagraphs \citep[e.g.,][]{brueckner_etal_1995} and the relatively new field of 
heliospheric imaging \citep[e.g.,][]{Eyles_2003,howard_etal_2008,deforest_etal_2022}.
Although Fresnel diffraction is important to certain parts of coronagraph design (specifically the occulter), 
ray optics are sufficient to design most of the instrument geometry, with recourse to wave optics
only for the detailed design of the occulter itself.

Externally occulted coronagraphs trade simplicity (of a direct occulter casting a shadow 
on the entrance to a camera or telescope) for fuzziness of the boundary of 
occultation on the image plane.  The fuzziness arises because external occulters at finite 
distance in front 
of the optics are imaged out of focus in the field of view.  That is because the corona itself is at
optically infinite distance, and therefore focusing the optics to image the corona necessarily defocuses 
the image of the much-closer occulter.

\begin{figure}
    \centering
    \includegraphics[height=6.5in]{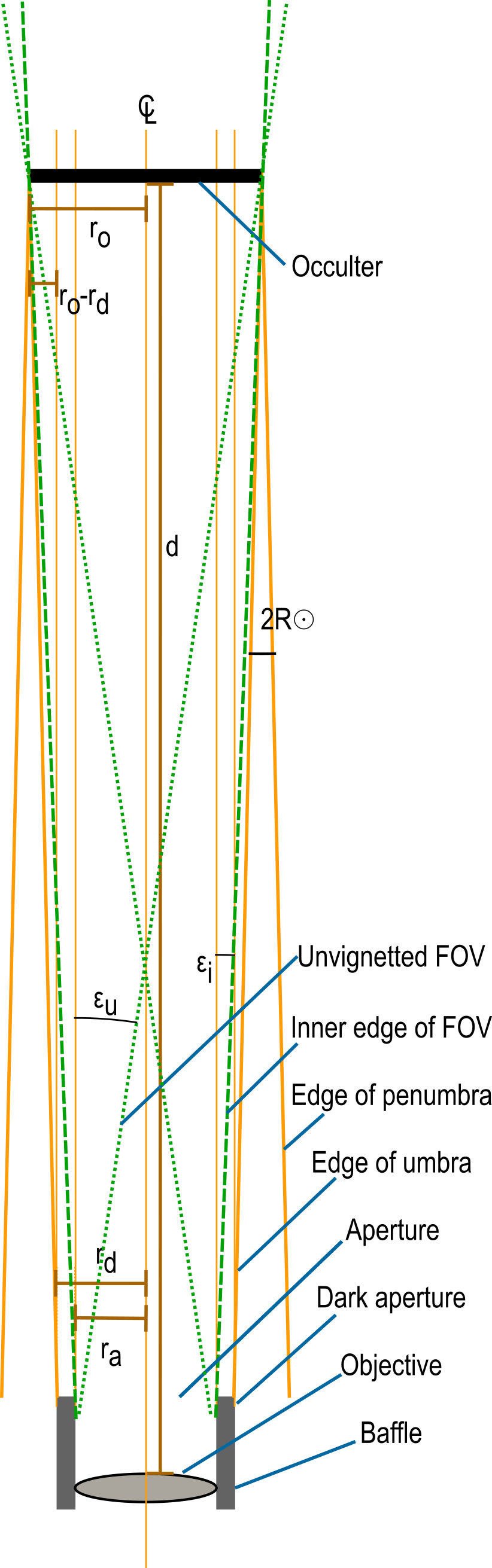}
    \caption{Geometry of a generic externally occulted coronagraph shows the related geometrical quantities
    used to specify even the simple CATEcor instrument.  The occulter is sized to completely shadow the aperture and 
    the dark baffle area.  The inner edge of the FOV, $\varepsilon_i$,
    is set by the angle between the edge of the occulter and the near edge of the aperture. 
    The innermost unvignetted portion of the FOV, $\varepsilon_u$, is set by the angle between the 
    edge of the occulter and the farthest edge of the aperture.  The umbra and penumbra extend inward and outward,
    respectively, from the edge of the occulter as shown. The spreading angle between the umbral and penumbral boundaries
    is the apparent solar diameter 2~$R_\odot$, and is exaggerated by a factor of 5 in this conceptual diagram.}
    \label{fig:ray-diagram}
\end{figure}

Figure~\ref{fig:ray-diagram}
illustrates the basic geometry of a generic dioptric externally occulted coronagraph, exaggerating the apparent size of the Sun 
by a factor of 5 for clarity.  
An occulter casts a shadow down the length of the instrument.  Penumbral and umbral edges formed
by the outer edge of the occulter are marked in orange.  The occulter is sized such that the edge of the
umbra lands outside a ``dark aperture'' that extends beyond the objective lens of the system.  
The principal design quantities are: $d$, the distance between the aperture and 
last effective plane of the occulter; 
$r_a$, the radius of the aperture; $r_d$, the radius of the dark zone behind the occulter; and $r_o$, the
radius of the final disk of the occulter itself. While instrument dimensions are lengths, the 
extent of the annular FOV is best described with solar elongation angles (apparent radial distances) $\varepsilon$.  
The inner edge of the FOV is set by the angle between the outermost edge of the occulter
and the nearest portion of the active aperture of the imaging optics, i.e.,
\begin{equation}
    \varepsilon_i =  \left( r_o - r_a \right) / d \label{eq:epsilon_i}\,,
\end{equation}
under the small-angle approximation that $\sin(\theta)=\theta$.  Meanwhile, the first non-vignetted elongation 
$\varepsilon_u$ is just
\begin{equation}
    \varepsilon_u = \left( 2r_o + 2r_a \right) / d \label{eq:epsilon_u}\,.
\end{equation}
In turn, the occulter size $r_o$ is determined by the dark radius $r_d$, the distance $d$, and the 
apparent size
of the Sun $R_\odot$.  For very long
instruments in which $r_d \ll r_o$, $r_a$ is negligible and $\varepsilon_u \ge 2~R_\odot$; for short
instruments such as CATEcor, $r_o \approx r_d$ and $\varepsilon_u \ge 4~R_\odot$.  CATEcor observations 
are vignetted throughout the anticipated effective FOV.

Note that $R_\odot$ \new{varies between the dates of perihelion and aphelion.}
\old{depends on the location of final deployment.}  
At the time \old{and 
location} of CATEcor's initial deployment, 2023 October\old{ on Earth}, 
$R_\odot$ is roughly 16{\arcmin} and therefore, 
\begin{equation}
    r_o \geq r_d + R_\odot d = r_d + 4.654\times 10^{-3} d \label{eq:occulter-size}\,.
\end{equation}
The actual diameter of the occulter is generally slightly larger than the equality of Equation~\ref{eq:occulter-size}, to allow for pointing error and/or a margin for Fresnel diffraction; in general, 
one can hold pointing margin in $r_d$ and use the equality rather than the inequality in Equation~\ref{eq:occulter-size}.

The inner edge of the FOV is the angle where the vignetting function reaches 0\%, i.e., the angle at which a single
ray can pass the occulter and enter the near edge of the objective.  It is thus determined by 
\begin{equation}
    \varepsilon_i = R_\odot + \frac{\left( r_d - r_a \right)}{d}\,,
\end{equation}
where again $R_\odot$ is the apparent solar radius (an angle) rather than the actual physical size of the Sun
(a length).  The narrower the occulted umbral buffer $r_d-r_a$ can be, the narrower the FOV.  
Further, for a given required umbral buffer $r_d-r_a$ around the aperture, larger/longer instruments 
can image closer to the Sun than smaller/shorter instruments.

The design drivers for the external shaded-truss concept are the structure of
the occulter and the design of the support. 

The occulter itself must combat 
Fresnel diffraction of sunlight around the occulting edge. Existing externally 
occulted coronagraphs use carefully aligned multi-disk designs with tight
positional and angular tolerances.  We designed an occulter form (Section~\ref{sec:occulter}) that greatly increases alignment tolerance while still 
reducing Fresnel diffraction (compared to a disk) and maintaining light weight.  

The shaded-truss design supports the occulter on a cantilever truss that 
must be stiff enough to maintain occulter 
alignment, while remaining fully within the umbral shadow of the occulter and also
obscuring as little of the FOV as practical.  We designed a 
simple hexapod truss built from narrow carbon fiber rods (Section~\ref{sec:shaded-truss}). 

\new{The length of the truss is only loosely constrained by the design.  In general, longer trusses perform better: the Fresnel diffraction brightness varies as $\sqrt{\lambda/L\theta}$, where $\lambda$ is wavelength, L is the length between the occulter and objective lens, and $\theta$ is bend angle (inner FOV edge).  The mechanical stiffness and buckling resistance of the truss constrains the length: occulter mass increases as $L^3$ and the first normal-mode frequency of the spring pendulum formed by the truss and occulter therefore decreases as $L^{-2}$.  We chose 75cm to place the first normal mode in the 10-20 Hz range with 2mm diameter carbon-fiber rods and an occulter under 50~g, and found that it met the Fresnel diffraction requirements for this particular instrument.} 

\subsection{Occulter design}\label{sec:occulter}

Occulters are limited in effectiveness by Fresnel diffraction, which 
allows direct sunlight to diffract around the sharp (projected) edge of the occulter and 
into the optics.  The effect of diffraction is reduced by larger instrument size or
by larger inner-edge elongation angles, and by multiple bends.  Fresnel scattering effects are in general 
complicated and require careful numerical analysis.  However, in simple geometries and approximations, 
Fresnel diffraction is tractable.  The straight razor-edge approximation to Fresnel diffraction can be performed
in 1-D and requires a cornu-spiral calculation \citep[e.g.,][\S 10.3.9]{hecht_zajac_1974}.  In the case of a single
plane-wave (collimated beam), the full integral
reduces to the Fresnel special functions $S$ and $C$, and may be written:
\begin{equation}
    I_{scatter} = \frac{I_B}{F^2} \left[ \left(C(\theta F) - \sqrt{\frac{\pi}{8}} \right)^2 + \left(S(\theta F) - \sqrt{\frac{\pi}{8}} \right)^2\right]\label{eq:Fresnel}\,,
\end{equation}
where $I_{scatter}$ is the intensity of light scattered around the occulter from the single collimated source to a 
detector, $\theta$ is the scattering angle around a nearly-straight section of occulter, $F\equiv\left(\pi d/ \lambda\right)^{0.5}$, $d$ is the length of
``throw'' from the occulter to the detector, and $\lambda$ is the wavelength of 
light being considered.  For polychromatic or white light, an integral over $\lambda$ is implied. The special functions $S$ and $C$ are defined by:
\begin{equation}
    S(\alpha)=\int_0^\alpha\sin\left(u^2\right)du\label{eq:S}
\end{equation}
and
\begin{equation}
    C(\alpha)=\int_0^\alpha\cos\left(u^2\right)du\label{eq:S2}\,.
\end{equation}
While Equation~\ref{eq:Fresnel} strictly applies to linear cases, it is appropriate and conservative for estimating
the stray light diffracted around the occulter provided that the ray-approximation impact parameter between a given ray 
at the aperture and the occulter itself is small compared to the radius of the occulter, i.e., for the observed ``bright ring'' of
Fresnel-scattered light observed near the occulter on the image plane \citep{howard_etal_2008}.

For $d=75$~cm, $\lambda$ over the range 450--650~nm, and an apparent occultation diameter of 1.4~R$_\odot$, 
averaging Equation~\ref{eq:Fresnel} across plane-wave contributions from the extended solar disk and across
wavelength yields a total scattering coefficient of $7\times 10^{-4}$, most of which is contained in the 
bright ring around the occulter.  Taking the in-lens scattering coefficient to be of order $10^{-2}$, 
which is conservative by roughly $10\times$ compared to typical values, this brightness corresponds to a
hazy background brightness of roughly $10^{-5}$~B$_\odot$ at 2~R$_\odot$ in the FOV.  During 
a 90\% annular solar eclipse, this brightness level is reduced by another order of magnitude, 
to $10^{-6}$~B$_\odot$, which is comparable to the expected sky brightness at 2~R$_\odot$ in Figure~\ref{fig:sky-brightness}.

Conventional occulter designs have evolved from single disks to multi-disk assemblies, so that multiple 
Fresnel scattering events are required for light to enter the instrument aperture 
\citep[e.g.,][]{howard_etal_2008,dudley2023}.  These assemblies can be difficult to manufacture and to 
align. One way to simplify alignment is to eliminate angular alignment altogether, by forming the occulter from a section of a sphere as in the spaceborne ASPIICS instrument \citep{zhukov2016} whose free-flying occulter is intended to operate hundreds of meters away from the aperture; this solution works well when the sphere's radius is long enough to enforce multiple Fresnel scattering events along the surface of the occulter, which is not the case for CATEcor.

We designed the CATEcor occulter to make use of a characteristic of
FDM 3D printing: printed objects are fabricated in layers,
which produces micro-ridges along the structure of the final object.  A gently curved
3D printed surface therefore approximates a more precisely machined set of edges
similar to the
edges of a multi-disk occulter. 
We initially considered a spherical envelope similar to that of ASPIICS because spherical
surfaces are simple to align.  But a spherical FDM occulter with diameter of a few cm would 
be sufficiently rounded that typically only one of the FDM ridges would interact with the light, 
and a spherical occulter would therefore behave optically like a single disk.

Multi-disk occulters in general have more gently curved envelopes than a similarly sized sphere.  The
ideal envelope is close to an ``ogive'': a figure of revolution of a large circle, about a chord near the 
perimeter of the circle. The ogive shape allows a constant angular offset between uniformly-spaced disks.
To improve the CATEcor occulter's effectiveness, we used an approximate ogive shape with sufficiently long
major diameter to allow multiple FDM ridges to interact with the light, thereby approximating the effect
of a series of very finely machined disks in a more conventional occulter.

\begin{figure}
    \centering
    \includegraphics[width=3.5in]{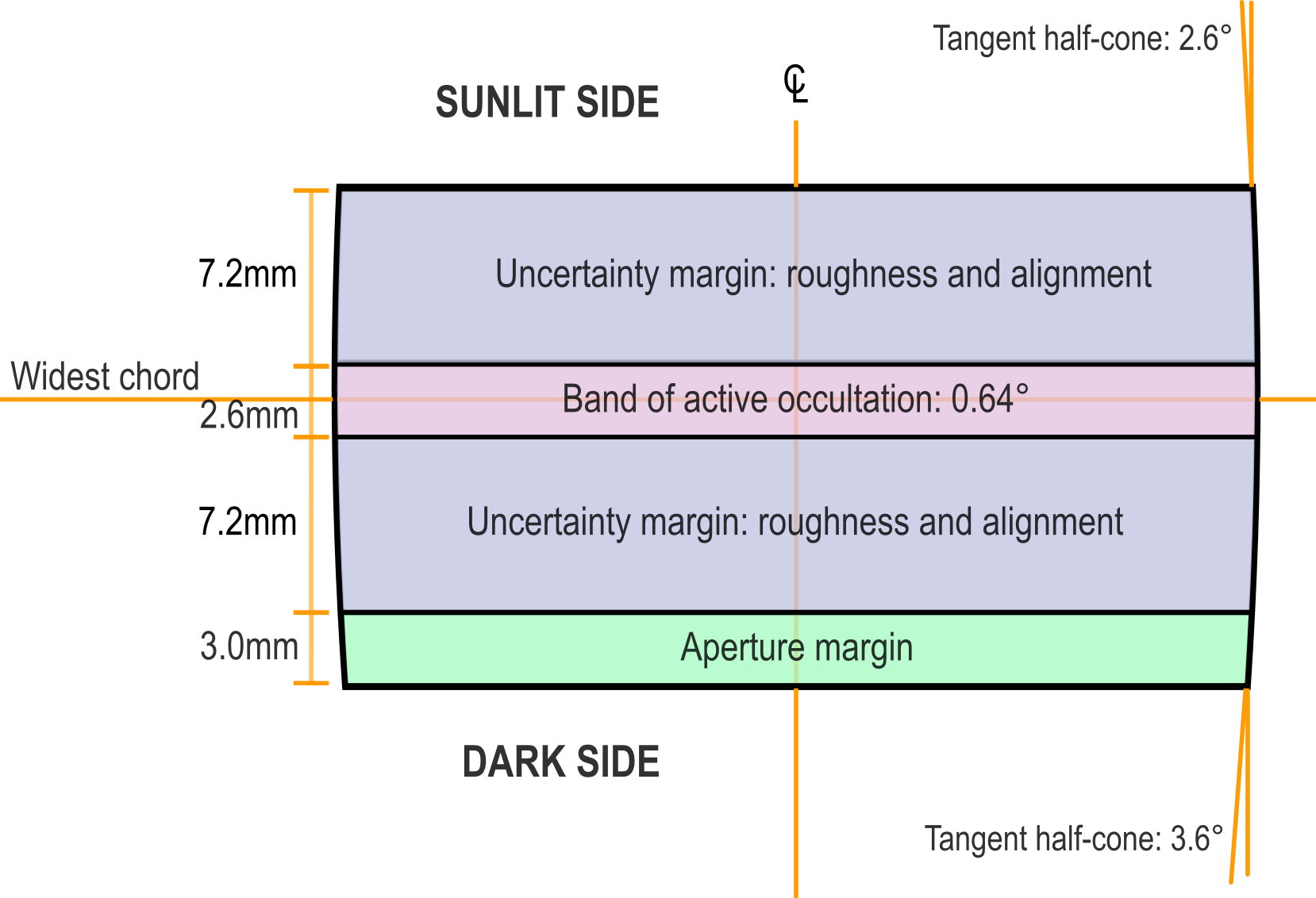}
    \caption{Conceptual cross-section of the CATEcor occulter shows the truncated ellipsoid, which approximates
    a shape with constant major radius of curvature (i.e. an ogive).  The occulter is a figure of revolution about the
    centerline.  The active occultation band obscures 2.4~$R_\odot$ of sky to block both the Sun's disk and the 0.4~$R_\odot$ design margin.  The additional thickness provides rigidity, mount holes for the occulted hexapod truss, and wide alignment tolerance of roughly $\pm 1${\degree}.  Not shown: center through-hole for alignment and six dark-side blind holes for the hexapod truss.}
    \label{fig:occulter}
\end{figure}

We approximated an ogive form with a prolate circular ellipsoid.  This choice was because ellipsoids are 
simple to create in 3D CAD programs, by stretching a spherical primitive shape.  We selected 
an ellipsoid with minor diameter
37~mm and major diameter 117~mm, i.e. stretched in the prolate direction by a factor of $\sqrt{10}$ compared
to a 37~mm diameter sphere.\new{ The 37~mm minor diameter arises from the inner FOV angle and the selected length of the truss suspending the occulter (Figure \ref{fig:ray-diagram}).}
\old{This }\new{The stretching }resulted in a local major radius of curvature of 185~mm at the equator of the ellipsoid, exactly 
10 times the minor radius of 18.5~mm. We retained only a small section of the ellipsoid, near its equator; this
yielded
an approximate truncated-ogive bowed cylinder shape as diagrammed in Figure~\ref{fig:occulter}.   This provides
for a significant length of interaction between the corrugated surface and 
near-tangent incident light from the solar disk, while also allowing some angular 
alignment tolerance for the assembly.  The truncation was slightly 
asymmetric about the equator, as shown in Figure~\ref{fig:occulter}, to provide a small amount of additional interaction surface to explore the interplay between stray light and aperture size.

The major radius of curvature of the CATEcor occulter is 185~mm at the widest point, and therefore 
the 0.4~$R_\odot$ angular separation between the solar umbra and the start of the FOV
imposes a separation of 340~$\mu$m between the point of tangency of rays from the solar limb and the 
closest point of tangency of rays that can enter the aperture of the instrument.  To enter the aperture, rays from the photosphere must thus traverse between 0.34~mm \old{(closest limb point)} and 2.06~mm \old{(farthest limb point)} of the occulter's outer envelope, to enter the aperture as part of the observed bright ring around the occulter.  \new{The shorter distance represents the smallest bending angle of solar rays; this is traversed by rays from the solar limb point closest to the edge of the FOV, to enter the outermost portion of the aperture.  The longer distance represents the largest bending angle, which is traversed by the ray from the solar limb point farthest from the edge of the FOV, to enter the outermost portion of the aperture. } Points not on the ultimate perimeter of the aperture are, of course, better shielded and require rays to curve around more of the occulter.  The 0.34--2.06~mm buffer zone on the curved envelope of the surface
reduces overall Fresnel scattering by a factor of 10--100 compared to a single
razor-sharp edge.  This establishes a zone from 0.86~mm above the widest point, to 1.81~mm below the
widest point, as the ``zone of occultation'' in ideal geometry.  However, other
uncertainties require an occulter thicker than the approximately 3~mm this would imply. 

Mounting and aligning the occulter is a significant challenge for an instrument that is designed
to be reproduced by students, and we therefore designed it with wide angular tolerance; this translates
to additional thickness beyond that required for the active occultation zone.  
Canting the occulter
by 1{\degree} moves the point of tangency up by 3.2~mm on one side 
and down by 3.2~mm on the other, while maintaining the overall 
occultation properties.
Further, the observed head positioning uncertainty of typical FDM 
3D-printed objects is 50--70~$\mu$m; that yields an uncertainty in the
occultation zone placement of an additional $\pm 4.0$~mm.  Thus the minimum height of the occulter is dominated by alignment uncertainties:  
a total of 7.2~mm are required on either side of the band of active occultation, for
combined fabrication and alignment tolerances.  

FDM printed objects are formed in layers, which form a corrugated surface with 
small bulges at the center of each layer and small canyons between the layers. 
The CATEcor occulters are printed at 50~$\mu$m layer thickness, ensuring at
least 50 layers across the zone of occultation, and a minimum of 6~layers 
for any one photospheric ray to bend around. Compared to a fully smooth polished
ogive surface, the layering yields a modicum of resistance to particulate 
contamination, by reducing the effect of invisibly small 50~$\mu$m-sized dust 
particles in the non-clean-room environment of a remote observing site. 

The CATEcor occulters are specifically designed for manipulation of the inner FOV, to
explore futher occultation if necessary.  Therefore they are extended 3~mm farther in the
direction of the instrument, providing meaningful additional deeper occultation (with a
wider occultation zone) out to 3.2~R$_\odot$.  The final design is thus a ``puck'' some 
20~mm tall: a truncated ellipsoid (approximating a truncated ogive) with minor radius 18.5~mm 
and major radius 185~mm, with the 
widest point \new{(widest cross section, at 37~mm diameter)} 1.5~mm above the centerline between top and bottom, and 8.5~mm behind the front 
(Sun-facing) surface.  As implemented with FDM printing, the puck is micro-corrugated at 
the 50~$\mu$m scale.

A through-hole at the center of the occulter permits alignment during assembly, and is
blocked in use by black adhesive tape.

The CATEcor occulters also contain 2.5~mm diameter blind holes to mount 2.0~mm o.d. 
truss rods during assembly. These holes occur in pairs on a radius of 14.5~mm from 
the centerline, and are tilted by 1.1{\degree}, to form the hexapod.  The holes are slightly 
oversized to avoid overconstraining the rods during assembly.

\subsection{Shaded-truss design\label{sec:shaded-truss}}

Externally occulted coronagraphs generally use a large baffled ``vestibule'' to 
control stray light, 
with an occulter supported by a rigid 
pylon \citep[e.g.,][]{brueckner_etal_1995}.  Instead, 
CATEcor supports its occulter with a carbon-fiber truss that is directly shaded by the 
occulter, simplifying stray light control by keeping the support structure out 
of direct sunlight.  
The design eschews the vestibule entirely.  The optical field of regard 
is limited by a
baffle 
mounted on the telescope tube, as diagrammed in Figure~\ref{fig:ray-diagram}.  
At a cost of 
increasing the complexity of the instrument's vignetting function, this support method 
provides simplicity, greatly eases alignment, and reduces total instrument mass and
complexity.

CATEcor uses the simplest possible fully-constrained truss: six rods mounted between 
two equilateral triangles of mount points, one on the 
occulter and one on the perimeter of the optical aperture.  
We chose carbon fiber for the rod material, for its ready commercial availability,  
high strength-to-weight ratio, and stiffness.  
The rod diameter is 2~mm, chosen to reduce vignetting while supporting a light occulter
at 75~cm distance from the mount.  The Euler buckling force limit for an unsupported 
75~cm long, 2~mm diameter carbon-fiber rod is 4~N, with
$10\times$ safety factor; CATEcor uses glue to fix the rod ends, providing angular 
support and increasing the safety factor by another factor of 4.  The occulter mass is
approximately 10~g, which imposes a bending force of 2--3~N on the truss when extended
horizontally, well within the capability of the 
six 75~cm rods composing the hexapod.


\begin{figure}
    \centering
    \includegraphics[width=4in]{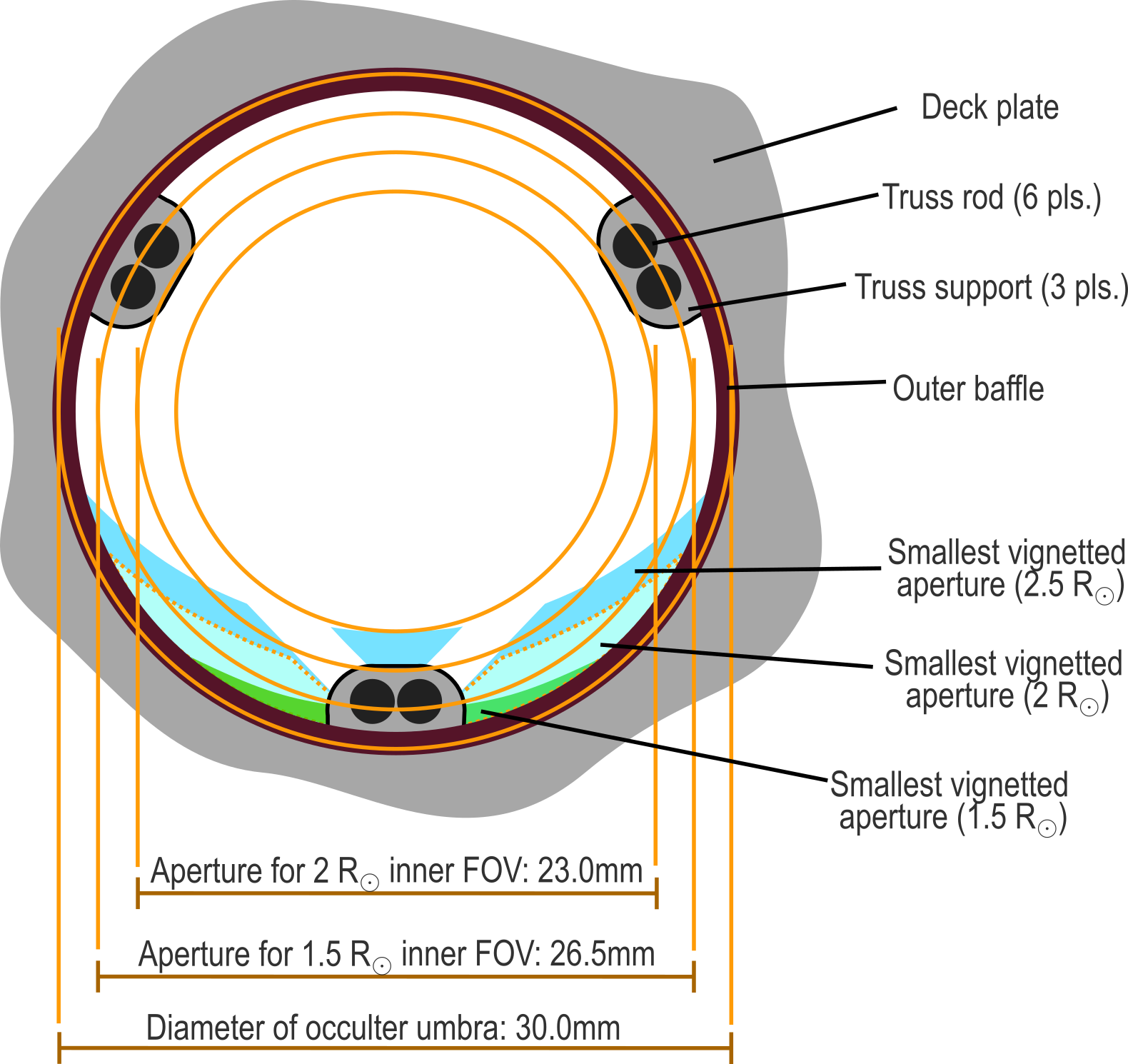}
    \caption{The CATEcor aperture, seen here from the point of view of the occulter, is sized to fit well inside the umbra of the occulter while still 
    supporting imaging at 1.5~R$_\odot$.  A surrounding deck encloses a precise iris aperture 
    (not shown).  With ideal pointing the umbra of the occulter forms a 30~mm diameter circle; the
    baffle is entirely inside the umbra.  The aperture is encroached upon by three small decks
    above the iris, supporting the six rods of the hexapod truss.  The decks partially vignette
    the innermost portion of the FOV.  The effective aperture is shown for three points on
    the sky, aligned with one of the hexapod supports: 1.5, 2, and 2.5~R$_\odot$ from Sun center.}
    \label{fig:aperture}
\end{figure}

\subsection{Aperture design} \label{sec:aperture}

To minimize the dark aperture region in Figure~\ref{fig:ray-diagram}, the CATEcor aperture 
is not a complete circle:  it is encroached upon by three decks supporting 
shaded hexapod feet, reducing the diameter of the required dark-shadow region.   The aperture and truss are 
guarded by a thin printed circular baffle (Figure~\ref{fig:ray-diagram}) whose leading 
edges are just inside the umbra of the occulter and just outside the aperture-plane triangle 
formed by the truss 
rods. The top few mm of the baffle are just 0.5~mm wide to separate the umbra from penumbra while
retaining as much open aperture as possible.

Figure~\ref{fig:aperture} shows the effective unvignetted aperture for three individual points on the 
sky:  the points at 1.5~R$_\odot$, 2.0~R$_\odot$, and 2.5~R$_\odot$, on the planned observing date of
14-Oct-2023 (apparent solar radius = 16{\arcmin}).  The top edge of each colored unvignetted aperture region
is defined by the occulter itself; the diagonal ``cutouts'' at 2~R$_\odot$ and 2.5~R$_\odot$ are from the
truss rods.

The instrument resolution is governed by Fraunhofer diffraction through the effective aperture, and is
anisotropic.  The radial diffraction limit at 1.5~R$_\odot$ is set by the 1.5~mm distance from top to bottom of 
the sliver of effective aperture, and is roughly 1.5{\arcmin}--2{\arcmin}, comparable to human visual acuity.  The 
tangential/lateral diffraction limit is roughly 0.3{\arcmin} at that distance from the Sun.  At 2~R$_\odot$, 
the radial diffraction limit is roughly 0.5{\arcmin}, and the lateral is under 0.1{\arcmin}.  Above 2~R$_\odot$, the effective resolution is likely to be limited by noise effects rather than seeing or diffraction
\citep{deforest_etal_2018}.

CATEcor includes an
optical-grade adjustable-iris aperture stop, located just behind the three hexapod decks.
The  stop is adjustable to explore the trade between inner FOV diameter and stray light with
smaller optical apertures.

\begin{figure}
    \centering
    \includegraphics[width=2.5in]{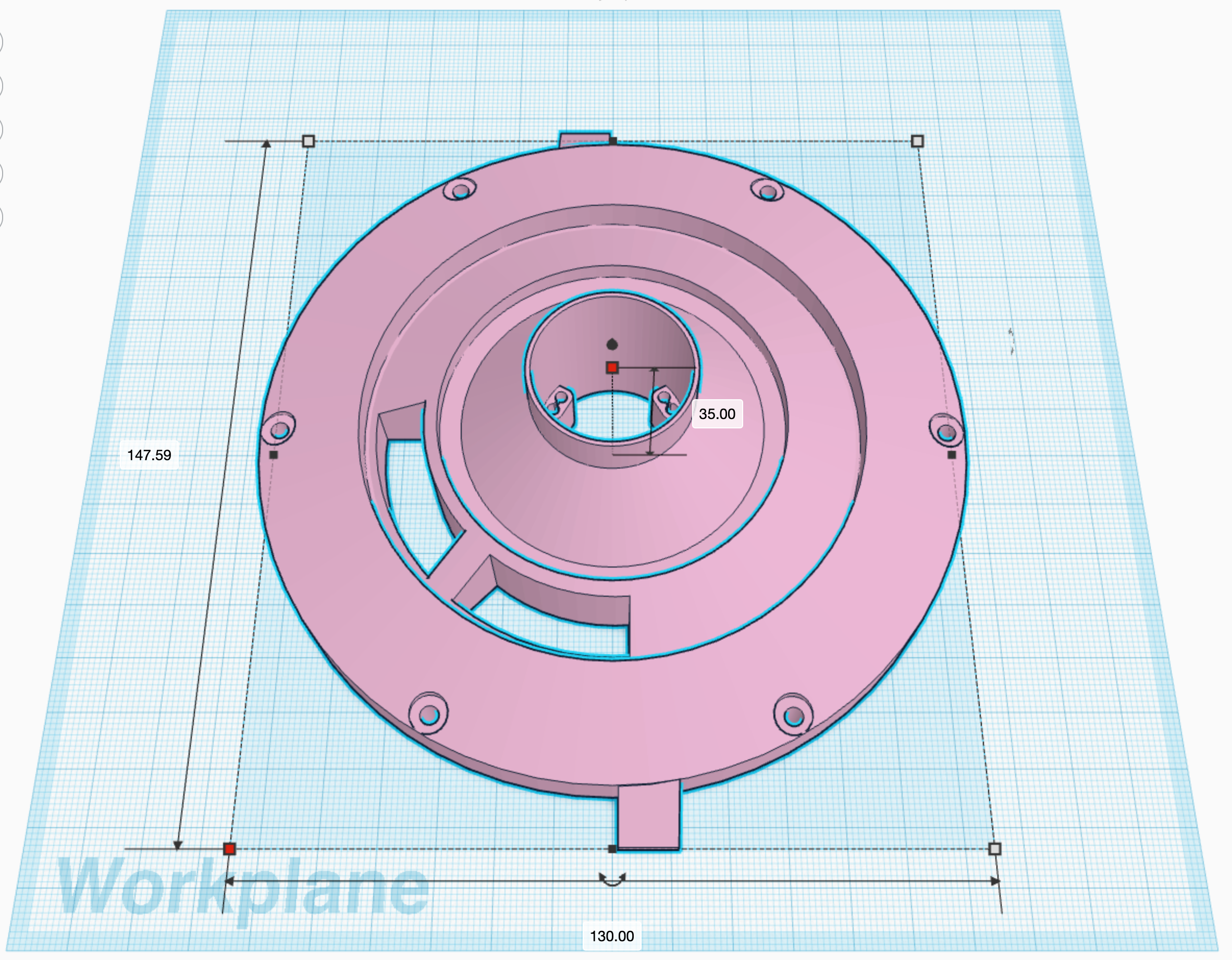}
    \caption{CATEcor aperture piece mounts to the CATEcor front plate, and is secured by six M3 bolts around the perimeter.  Lateral stability is ensured by positive lock between the tabs on the front plate and matching recess holes on the underside of the aperture piece.  1/4 of the perimeter has a slot to allow adjustment of the adjustable-iris aperture stop in the interior.  Small finger tabs on the perimeter help to mate/demate the press-fit features with the front plate. Two of the three hexapod mounting decks can be seen at the base of the central aperture.}
    \label{fig:aperture-drawing}
\end{figure}

The 28~mm diameter primary aperture, hexapod supporting decks, and tube/rim baffle are 3D printed as a 
single piece (Figure~\ref{fig:aperture-drawing}).  The aperture piece supports the hexapod
and is supported by a front plate located behind it.  Each hexapod deck has two 10~mm deep blind holes, 2.5~mm in diameter, canted at 
0.6{\degree}, to support the ends of two of the 2~mm diameter hexapod rods.  The holes are
slightly oversized compared to the rods, to provide alignment play and to allow room for 
glue to bond the rods rigidly to the aperture piece.

The aperture piece is fixed precisely relative to a supporting
front plate, by six radially-aligned rectangular alignment holes in the bottom surface.  These mate
with rectangular-extrusion alignment tabs in the front plate.  The pieces press-fit together, and small
finger tabs are provided to mate/demate the pieces.  The aperture piece is secured by six M3 fasteners which 
extend, via through-holes that penetrate the aperture piece and front plate, into captive square nuts in
a supporting telescope tube.

A slot in the aperture piece gives tool access to an adjustment lever on an adjustable-iris
aperture stop underneath.  The slot is 10~mm wide, too narrow for fingers but wide enough to
reach in with a small screwdriver or Allen key.  The slot is interrupted by a plastic 
beam to maintain
stiffness of the entire part. The top surface is radially beveled to reduce glinting stray
light after surface treatment (painting).  The 
slot for adjustable-iris control acts as a light trap in use, and does not noticeably increase stray 
light or glint.

\subsection{Front plate design\label{sec:front-plate}}

The aperture piece rests on a front plate interface that accepts and supports an adjustable-iris 
aperture stop assembly, and in turn rests on a telescope adapter tube.  The aperture stop assembly is bolted in place 
with three M6 through-bolts that mate with captive 
hex nuts on the underside of the front plate. The caps of the bolts recess into holes in the aperture 
piece.  The bolt holes are slightly oversized to allow adjustment of the aperture stop's lateral 
position during assembly, before the bolts are torqued down.  Positional alignment is maintained between
the mounted aperture piece and the bolted-on aperture stop, via six extruded alignment features that
mate with alignment holes on the underside of the aperture piece.  A circular 
mounting ring, together with a corresponding groove on the underside of the aperture piece, forms a four-bounce ``optical maze'' to prevent stray light entering the dark space behind the aperture.

\subsection{Telescope tube extension design\label{sec:telescope-tube}}

The telescope tube extension is fabricated with 10~mm thick walls for
stiffness, and is secured to the CATE24 telescope tube by six nylon 
M8 bolts forming a dual triangular friction mount.  The nylon M8 bolts
are retained by hex nuts held captive by interior hexagonal holes.  
The front plate and aperture piece are secured to the tube 
with six steel M3 bolts that engage with six captive square 
nuts near the top of the piece.

\section{Fabrication and Integration}\label{sec:integration}

All 3D printed parts of CATEcor were printed on a hobbyist 3D FDM printer 
(PRUSA Mk 3) in black polyethylene terephthalate glycol (PETG) plastic.  The 
FDM settings depended on the part.  The telescope tube and front plate were not 
required to have particularly precise 
shape and we printed them at high speed with 200~$\mu$m layer thickness and 15\% 
infill.  For the aperture piece we used ``precision''
(lower speed) extrusion settings, with 150~$\mu$m layer thickness, with 30\% infill and
doubled perimeter wall thickness and top/bottom surface layer count, for rigidity.  
The occulter 
required highly precise form and was iterated several times to optimize 
printer-specific parameters for the cleanest print. 
We printed it with 50~$\mu$m layer thickness, with external perimeters deposited first
to provide the cleanest possible exterior
shape, and 30\% infill.

\begin{figure}
    \centering
    \includegraphics[width=3.5in]{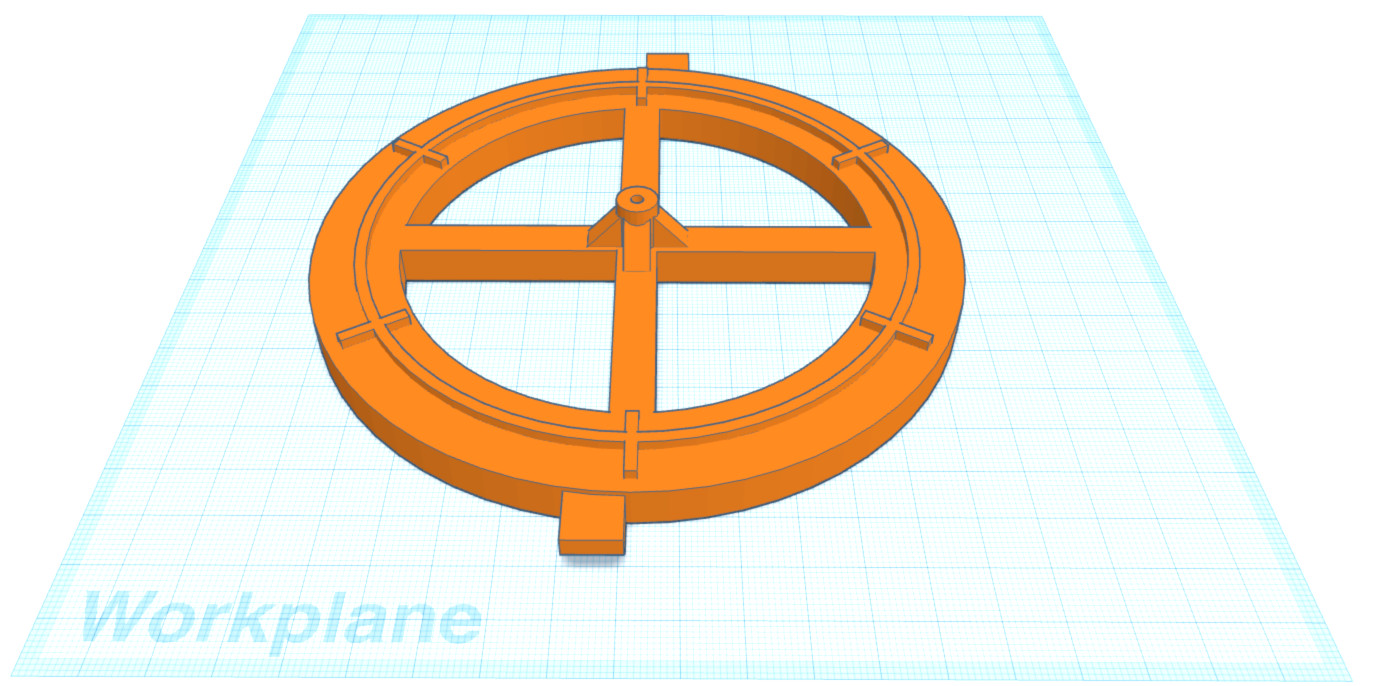}
    \caption{A jig piece used to align the occulter assembly matches the shape of the CATEcor front plate, with rigid support for a seventh alignment rod at the center of the assembly.  The central gusseted tube fits into the central hole of the CATEcor aperture piece, to align the aperture piece and occulter while the hexapod
    truss undergoes assembly and gluing.}
    \label{fig:CATEcor-jig}
\end{figure}
 
The trickiest and longest integration step was assembling the front-end assembly
comprising the occulter, hexapod rod truss, and aperture piece. We describe that process here, to illustrate its simplicity and the lack of any special tools
required.

To align the occulter
we printed a separate jig piece (Figure~\ref{fig:CATEcor-jig}) that accepted a
seventh 2.0~mm diameter rod.  We placed the jig directly under an alignment hole drilled
into a ``2$\times$2'' wooden alignment beam that was rigidly mounted 90~cm above the work 
surface. We aligned the jig piece using a plumb bob, then marked the work surface
at the location of
the jig piece so that the jig piece could be moved and replaced at the same location.

We secured the rods to the occulter and aperture piece with household 2-hour cure epoxy.
After mixing the epoxy, we placed the occulter, upside down, on a clean work surface, 
and used a bit of rod scrap to pack the six hexapod mount holes in the occulter
approximately 25\% full of epoxy.  We wiped up excess epoxy using a paper towel, wiping
radially inward from the edge of the occulter to avoid smearing any epoxy on the active
surface.

We mounted an aperture piece on the jig, then threaded the occulter, sunlit side up, 
onto an uncut (120~cm long) 2~mm rod, threaded the top of the rod upward through the hole
in the alignment beam, then threaded the bottom of the rod downward through the aperture 
piece into the central tube of the jig.  Finally, we placed the jig inside the alignment
marks on the work surface.  This provided lateral positional alignment of approximately 
$\pm$2~mm between the occulter and aperture piece, equivalent to angular alignment of 
$\pm$10{\arcmin}.  The occulter central hole fit tightly on the central alignment rod, 
affording approximately $\pm$20{\arcmin} of angular tolerance for an overall 
$\pm$0.5{\degree} 
alignment precision compared to the $\pm$1{\degree} design tolerance.

Once the jig, aperture piece, alignment rod, and occulter were placed under the 
alignment beam, we assembled the hexapod. For each rod, we wiped the ends with an
isopropanol-wetted paper towel, waited a few seconds for the ends to dry, then dipped
one end approximately 1~cm deep in the epoxy, wiped it on the mixing container to remove
excess, carefully inserted the glue-covered end into one of the mount holes in the 
aperture piece, 
and finally bowed the rod to insert the clean end into the corresponding pre-packed 
mounting hole in 
the occulter.  During the insertion, we secured the occulter by grasping it between 
thumb and forefinger, by the sunlit and shaded flat surface (not the active surface).
For each rod, we examined the hole ends for excess epoxy and wiped any 
excess using the end of a 1.5~mm flat-blade screwdriver.  We assembled the rods in 
circular order around the hexagonal assembly.  When all rods were inserted at both ends,
we visually inspected the occulter for placement, and -- holding the rods and not the
occulter -- we rotated the occulter roughly 45{\degree} to either side of its ideal 
alignment, to further spread the epoxy in the holes, before aligning the occulter
rotation angle.

To set the rotation angle we sighted through the truss, 
standing along one of the six mirror symmetry planes of the truss assembly.  Pairs
of rods form vee shapes at the bottom of the truss, and complementary pairs form 
inverted-vee shapes at the top of the truss.  Standing so that one vee on the front side
bottom lines up with the central rod should also cause the complementary inverted-vee on
the rear side top to line up with the central rod.  We twisted the occulter into best
visual alignment by sighting along one of the three major symmetry planes and verified
alignment by sighting along the other two. 

We allowed the epoxy to cure for 8~hours, then removed the central rod from the jig, and
removed the aperture-truss-occulter assembly.  We verified rigidity by plucking the 
truss while holding the aperture piece firmly against the work surface, to observe the
low oscillation modes and verify the low-amplitude fundamental to be at or above 10~Hz. 
Because the hexapod is the minimum complete support system, even one loose glue joint has
a large effect on the fundamental mode frequency and is obvious.

Based on initial testing (Section~\ref{sec:initial-testing}), we treated the
surface to reduce glint.  We applied a light coat of commercial Krylon
ultra-flat camouflage black spray paint to the entire occulter assembly, holding
the spray can some 8~inches from the occulter and lightly ``spritzing'' the 
paint.  Similar treatment was applied down the length of the truss and in the
interior of the aperture assembly, to prevent glint from the finished plastic.
Particularly on the occulter itself, and also on the truss, we did not attempt
full coverage, just a 
single very light coat.  
This was sufficient to intercept any glint coming along
the occulter at grazing incidence.  We tested both coated and uncoated
assemblies and found the stray light from uncoated assemblies to be reduced
by 3$\times$ or more in the coated assembly (Section~\ref{sec:initial-testing}).

\begin{figure}
    \centering
    \includegraphics[width=4in]{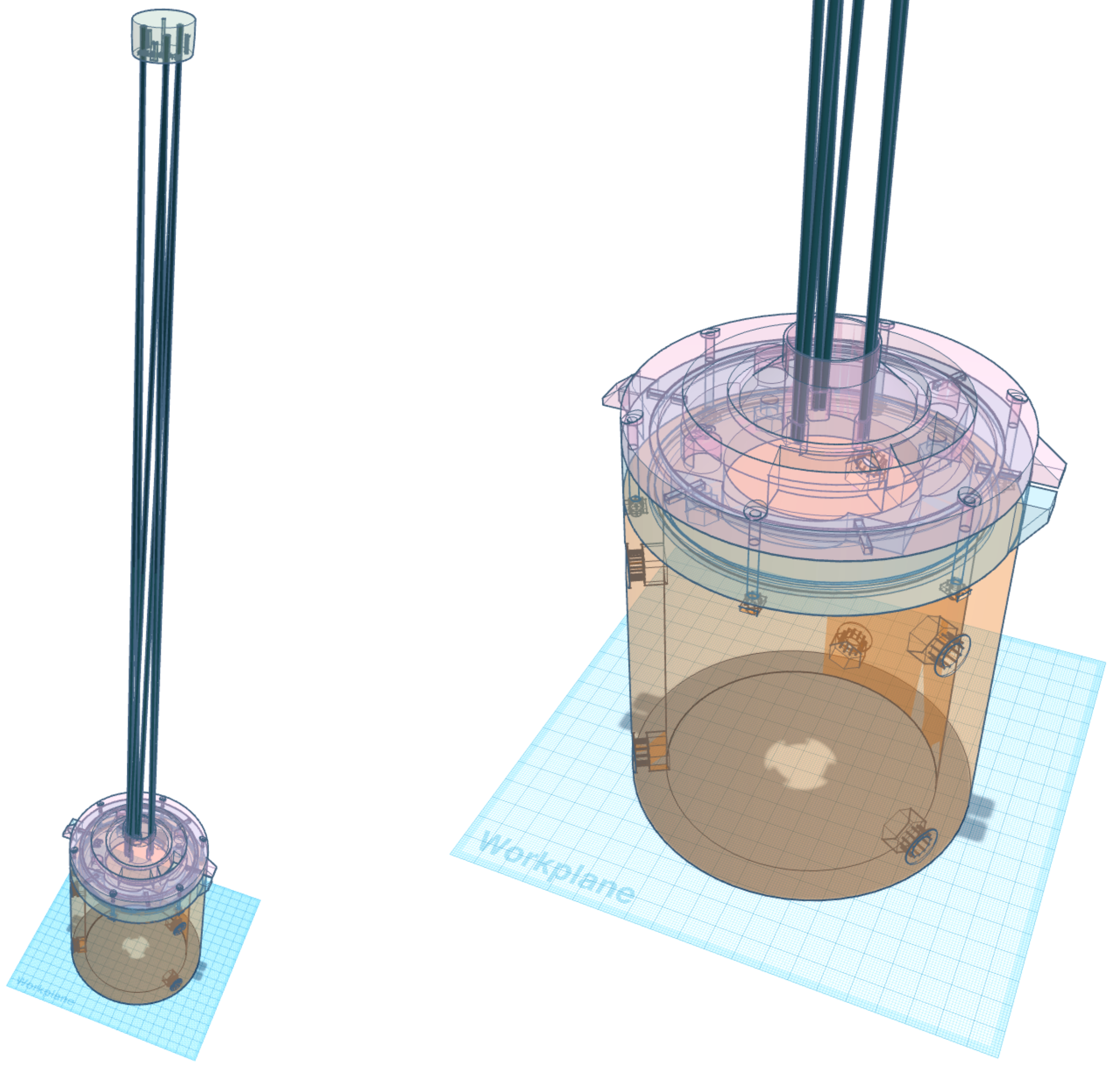}
    \caption{CATEcor adapter assembly drawing shows how the various parts, including occulter, fit together.  All parts are 3D printed using fusion deposition modeling at various layer widths, except for the truss rods -- which are 2~mm diameter pultruded carbon fiber, cut to 76~cm length. }
    \label{fig:adapter-assembly}
\end{figure}

Assembling the rest of CATEcor is more straightforward.  We fitted M8 nylon
screws with a 
bolt and washer, then threaded them through the holes in the telescope 
extension tube and secured them with captive nylon nuts in the interior 
hexagonal holes.  We fitted a commercial iris 
into the front plate and bolted it in with M6 bolts and captive nuts. 
We placed the front 
plate on the telescope interface tube alignment ring.  We placed the aperture/occulter
assembly on the front plate, so that the ejection tabs on the perimeter 
approximately lined up, then secured the entire assembly using six M3 bolts through
the six perimeter holes in the aperture assembly, into captive square nuts in the 
telescope extension tube. This resulted in the full assembly shown in Figure \ref{fig:adapter-assembly}.

To observe, we integrated the CATEcor assembly to a CATE24 telescope, on-site at an observing location.  We developed a procedure to avoid direct solar exposure through the unprotected
telescope onto the detector.
We first aligned the telescope and aimed it at the Sun, tracking solar rotation, with a solar filter
on the objective lens.  One person held a shadow mask above the telescope, casting a shadow 
onto the objective, while another person removed the solar filter and replaced it with a CATEcor assembly.
We stopped down the iris aperture, and iterated a process of briefly removing the shadow mask to observe
the shadow of the occulter, then adjusting the nylon set screws to bring the shadow closer to being centered
on the aperture and checking alignment again.  After 3--4 iterations the occulter shadow was centered over the 
aperture and we removed the shadow mask entirely.  During observation, we observed the aperture periodically and
used the telescope pointing controls to re-center the solar image and shadow as needed.

\section{Initial testing}\label{sec:initial-testing}

We constructed an engineering unit of the CATEcor coronagraph front-end assembly,
and performed simple testing on it including deployment on a CATE24 telescope on a 
sunny day.  

\begin{figure}
    \centering
    \includegraphics[width=4.8in]{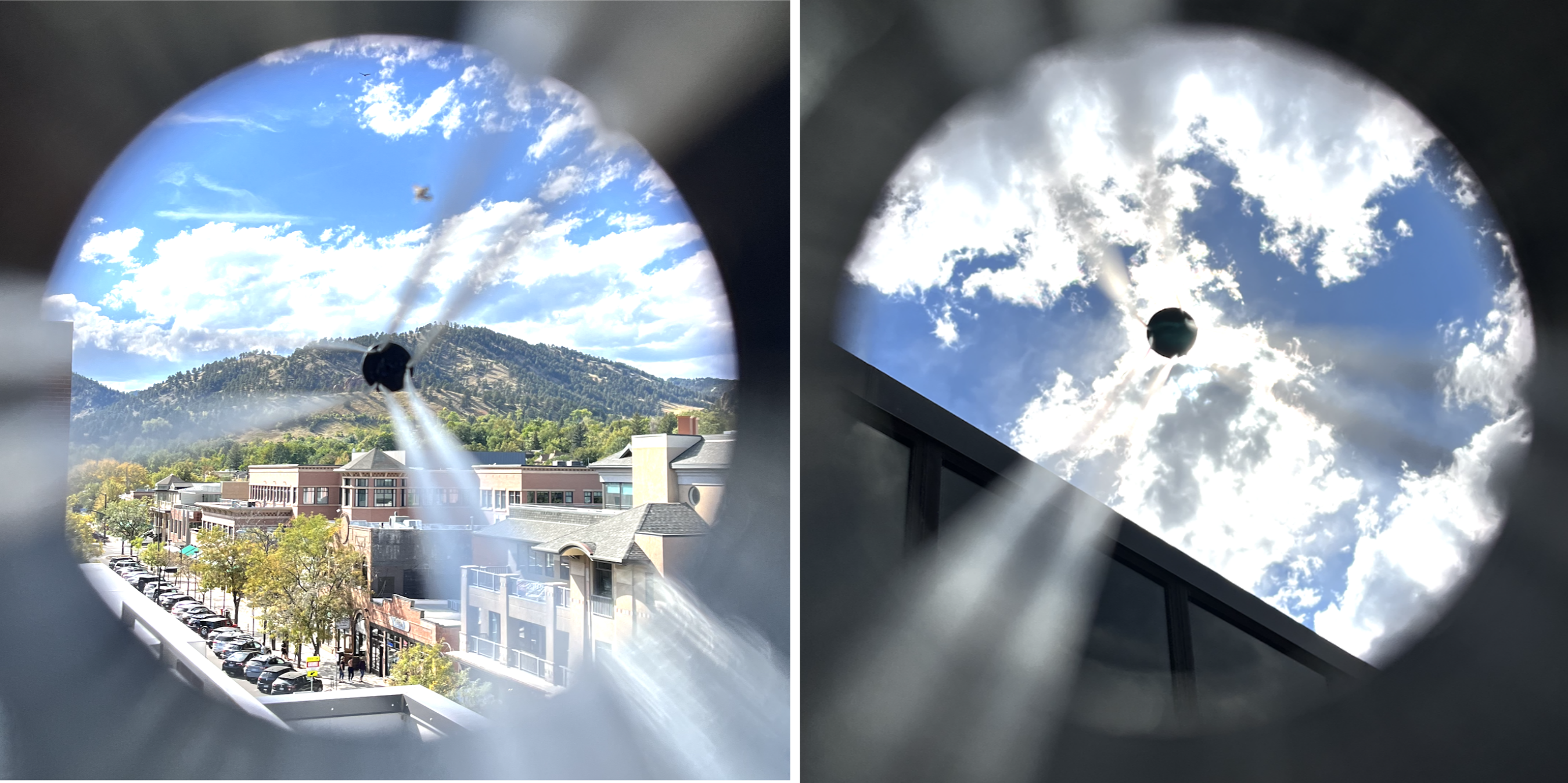}
    \caption{Initial test images taken through the CATEcor engineering unit reveal the geometry of the instrument front-end.  Left: the Colorado front range above Boulder, CO; Right: the Sun with light clouds.  These images were collected with a cell phone camera held at the aperture at the rear of the assembly.  The truss is more strongly visible and the occulter appears slightly larger than in final images, because the cell phone effective aperture is much smaller than the full available aperture at the rear of the assembly.}
    \label{fig:test-images}
\end{figure}

The aperture-truss-occulter assembly, fully assembled, has a mass well under 1~kg 
and is readily manipulated by hand.  It is easy to sight through the aperture 
and occulter itself directly with one's eye or with a small camera.  Figure~\ref{fig:test-images} is the result of a crude initial 
optical test: a photograph of the Colorado Front Range above Boulder, Colorado
taken through a cell phone camera held at the aperture of the assembly; and an exposure of the Sun through light clouds taken in the same circumstance.

The truss structure is more strongly visible in Figure~\ref{fig:test-images} than in scientific images through a CATE24 telescope, because the cell phone aperture is much smaller than the available aperture at the rear of the occulter assembly.  The truss takes the appearance of three pairs of ``parallel'' rods extending to the occulter.  The ``parallel'' rods are actually diverging from the aperture side of the hexapod to the occulter, and the widely spaced rods are converging toward single mount points on the occulter.  

\begin{figure}
    \centering
    \includegraphics[width=4.8in]{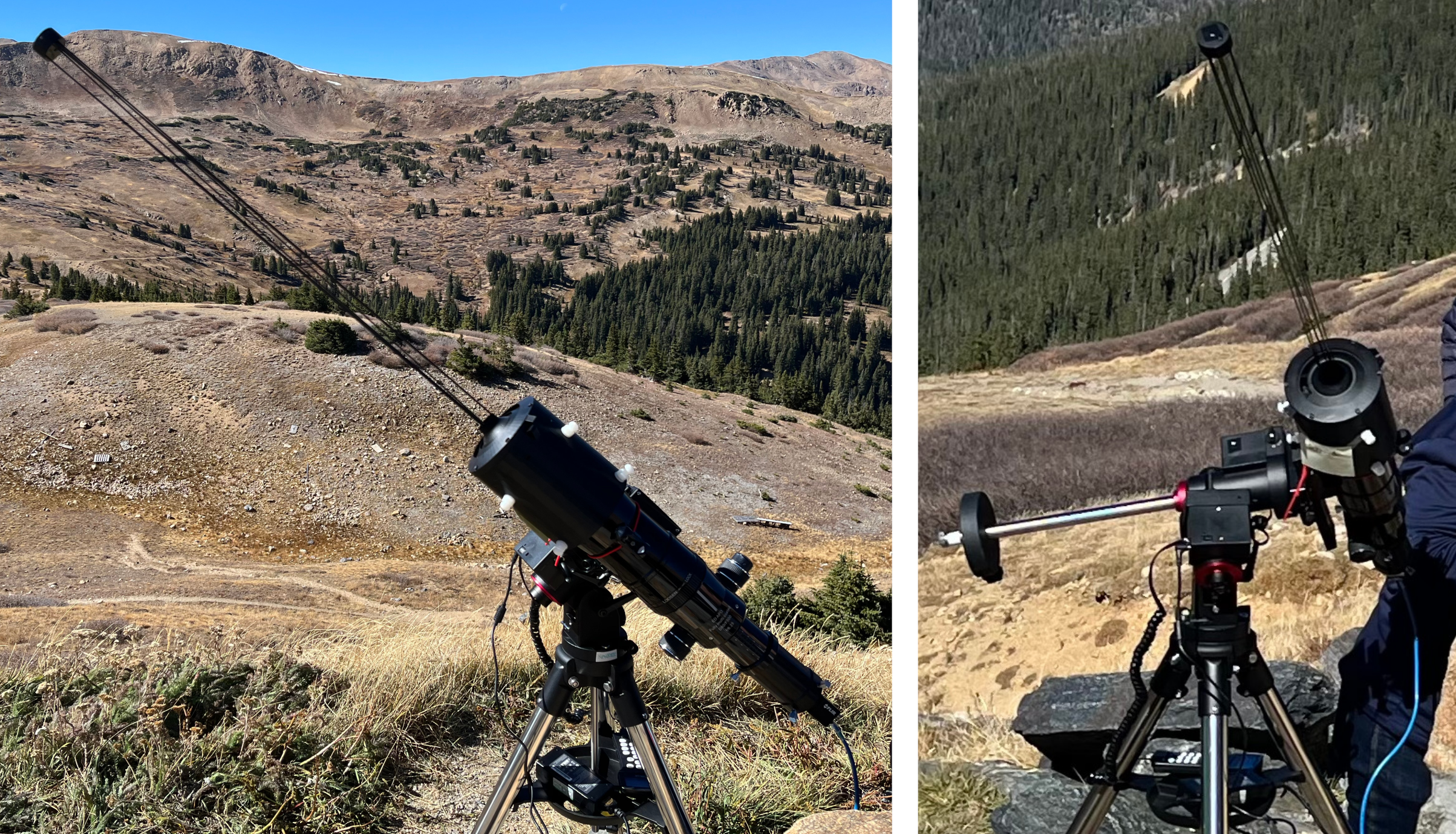}
    \caption{Initial deployment of the CATEcor at Loveland Pass, Colorado, on 6-Oct-2023, helped establish alignment procedures, demonstrated dark occultation, and revealed some glint around the uncoated occulter. The side view shows the 75~cm distance between the rear of the occulter and the interior of the aperture piece.  The front view shows the apparent extreme darkness of the instrument aperture, which is bathed in deep umbral shadow from the occulter.}
    \label{fig:deployed}
\end{figure}

After full integration, we deployed CATEcor on a 
clear sunny day at an altitude of 12,000 ft. at Loveland Pass, Colorado, on 
6-Oct-2023, in an interval surrounding noon (roughly 12pm--3pm MDT; noon occured at 12:54pm).  
The deployed instrument is shown in Figure \ref{fig:deployed}.
Despite a low cloud layer to the north and 
northeast of the observing
site, the sky color at the pass was dark cerulean to cobalt blue \new{with minimal white (Mie-scattered) halo around the Sun}, reflecting 
good ``coronal sky''
conditions.  At the pass we observed some light specks in the air, which we
speculated to be high-flying pollen from aspens or high-altitude grasses.  
Wind levels varied from 5 to 15 knots through the course of the observation, 
resulting in visible wind shake of the occulter shadow when the occulter was
aligned on the Sun.  The wind did not visibly shake the telescope itself \citep[see][for addition\new{al} description of our initial field test]{Seaton_etal_2024}.  

\begin{figure}
    \centering
    \includegraphics[width=4.8in]{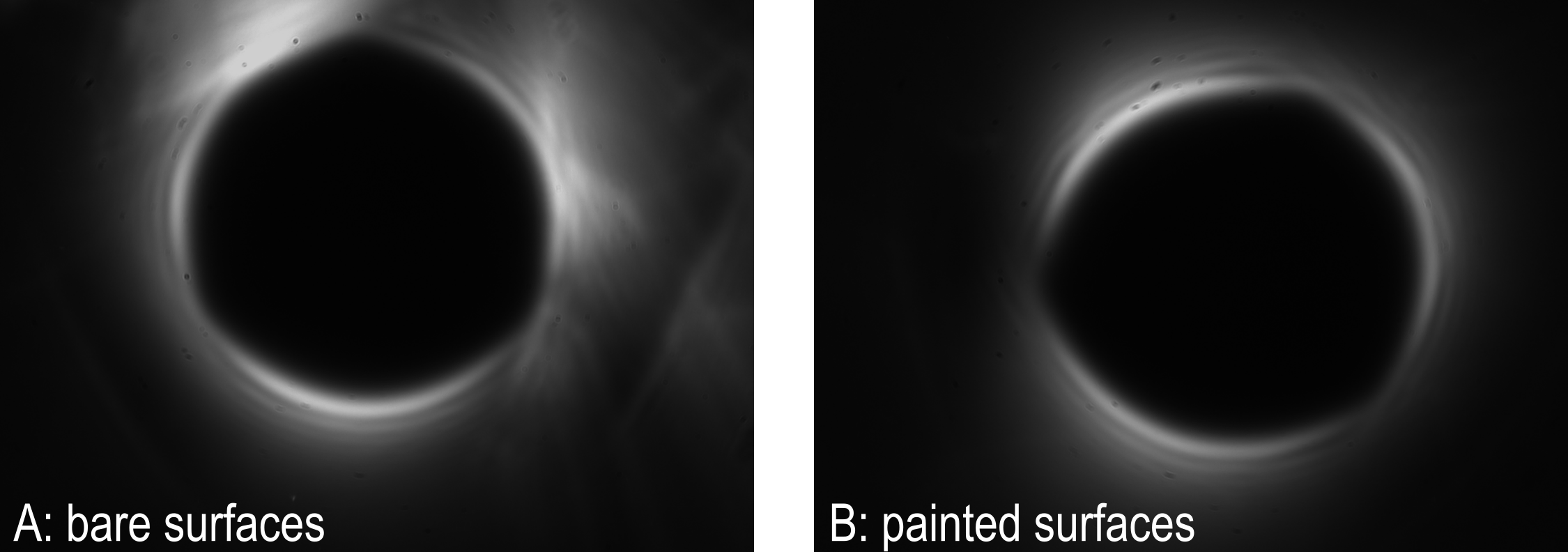}
    \caption{Bare/painted either-or test images with two copies of the CATEcor occulter show the importance of surface treatment to prevent specular reflection and glint.  Both images were collected with a single CATE24
    telescope and two separate Sun-pointed CATEcor occulters: one bare, and one painted.  These images are 
    from Loveland
    Pass, Colorado on 13-October-2023, and have the same exposure time and scaling.  Panel (A) shows the diffraction/stray-light pattern with bare surfaces.  Panel (B) shows the diffraction/stray-light pattern with painted surfaces.  Angular glint features and overall background
    are greatly reduced with the painted occulter.}
    \label{fig:paint}
\end{figure}

Initial images from 6-Oct-2023 were sufficiently
dark that we proceeded to manufacture and integrate several more occulter assemblies; but the images did include features
that we surmised to be glint.  
To half of the new occulters, we applied a flat surface treatment as described
in Section~\ref{sec:integration}; on the other half we left the occulter and truss surfaces un-coated.  
We performed 
either/or tests in similar conditions by swapping out and re-aligning the occulter assembly from a fully
assembled CATEcor at each of two observing sites for the 14-Oct-2023 annular/partial eclipse, on 13-Oct-2023 (the
day before the eclipse itself).  The two observing sites produced comparable results and revealed that 
flat black paint reduces glint without appreciably affecting the diffraction pattern around the occulter.  
Figure~\ref{fig:paint}
shows typical stray light images from the uncoated and coated occulter assemblies in full Sun.  

Figure~\ref{fig:paint}, Panel B, shows a typical stray light pattern from a coated CATEcor occulter assembly.
The familiar diffraction ``bright ring'' is visible, and is conjugate to an Arago spot \citep{Arago_1819} formed 
by the occulter at the entrance aperture of the telescope.  The ring is interrupted in three places by vignetting
due to the truss holding the occulter. The bright ring has an inner radius of approximately 1.3~R$_\odot$,
and extends to 5~subsidiary peaks extending to 1.5~R$_\odot$.  The extent of the subsidiary peaks is a measure of
precision in the figure of the occulter.  It is a sign both that the signal at this dynamic-range setting is 
dominated by diffracted light as expected, and also that the occulter and truss work as designed: while present in
the FOV, the shaded truss does not significantly contribute to the stray light pattern observed in the
final CATEcor coronagraph images.

We used painted CATEcor occulters on CATE24 telescopes at both Loveland Pass, Colorado and Sandia Peak, New Mexico
to observe the partial or annular (respectively) eclipse of 14-Oct-2023.  Results of that experiment, including initial detection of the 
solar corona out to 2~R$_\odot$ from Sun center, are reported in \citet{Seaton_etal_2024}.

\section{Discussion}\label{sec:discussion}

The CATEcor coronagraph, as built, is best described as a proof-of-concept instrument.  We have conceived,
designed, produced, and tested a novel class of instrument, a shaded-truss externally occulted coronagraph,
that takes advantage of the umbral shadow of the external occulter itself to reduce (and effectively eliminate)
stray light from the support structure.

\subsection{Design advantages}

The shaded-truss concept has several advantages in the coronagraph design space.  By separating the functions of
stray light control and occulter support, shaded-truss designs greatly reduce the bulk of the coronagraph 
front-end.  In the CATEcor design, only a very small baffle isolates the aperture itself from the bright 
sunlight impinging on the instrument, essentially reducing the front-end stray light control structure to a
minimalist support and a very lightweight occulting body.  Further, because the front-end support is \new{very light}\old{as light
as possible}, it is feasible to suspend an external occulter at surprisingly large distances from the aperture,
improving performance compared to designs with a stray-light control structure (i.e., vestibule)
surrounding the occulter and a conventional support pylon.  

These advantages of the shaded-truss approach are sufficiently great that we were able to conceive, design, manufacture, 
integrate, and test a coronagraph front-end in under six weeks, using freeware CAD tools, hobbyist parts, and a consumer-grade 
FDM 3D printer.  Despite the preposterously simple design, fabrication, and 
integration approach -- which are accessible to any hobbyist with access to a ``makerspace'' workshop and a
good hardware store -- the resulting coronagraph is demonstrated to be cleanly diffraction limited and readily deployable in 
combination with common amateur-astronomy equipment.  This, in turn, implies that shaded-truss coronagraphs have 
potential to be important tools either for educational and amateur coronal viewing, or for quantitative scientific work.  We also demonstrated
this particular design by observing the Sun's corona at the annular (partial) eclipse of 14-Oct-2023 \citep{Seaton_etal_2024}.

Compared to conventional externally occulted designs, shaded-truss coronagraphs can bring the occulter farther
from the aperture, reducing the limitations of external occultation.  The ideal external occulter is 
infinitely far from the aperture and infinitely large, as exemplified by the Moon during a total solar eclipse.
Existing externally occulted designs use separation distances well under a meter, largely because it is infeasible,
in a deployable or spaceborne instrument, to build a large combined structural and stray-light-controlling 
structure in front of the optics.  This limits the sharpness of the partially-vignetted zone around the Sun, 
and therefore the inner radius of the FOV of externally occulted coronagraphs.  
By permitting a long cantilever distance between the aperture and occulter, 
shaded-truss designs better approximate the ideal conditions of a total solar eclipse, lessening the primary
disadvantage of external occultation.

Conventional externally occulted coronagraphs have strictly limited fields of view, which are determined by the
diameter of the aperture in the leading structure.  By divorcing the functions of structure and stray-light control, a shaded-truss coronagraph affords a much
broader FOV.  The CATEcor outer FOV was limited by the CATE24 telescope itself at a few solar radii;
but the occulting assembly, which was not particularly optimized for FOV, could nevertheless admit images beyond 
40~R$_\odot$
in all directions (Figure~\ref{fig:test-images}).  This permits a broader design space for future instruments
that could in principle cover from a few tenths of a solar radius to many degrees from the Sun in a single 
field of view.

\subsection{Lessons learned\label{sec:lessons}}

Several immediate design improvements are apparent from this proof-of-concept study. 

The occulter itself uses the corrugation inherent to 3D printing to force multiple diffractive scatters for
light to enter the primary aperture.  We found, on deployment, that this fine corrugation was not sufficiently
deep to prevent dust and other forms of contamination from ``spoiling'' the cleanliness of the occulter's active
surface.  On deployment, we noticed that small bright quasi-glints could be seen around the perimeter of the
occulter.  Close inspection showed that these glints were dust particles, fibers, and other contaminants that landed
on the surface during observation and extended into the bright sunlight around the occulter itself. Designing and
printing a more deeply, explicitly corrugated surface would greatly reduce the effect of this type of contamination,
while not affecting strongly the other properties of the occulter.  Likewise, blowing off dust and lint with filtered, clean air or nitrogen just before acquiring data may improve future observations.

Aligning the occulter and telescope to the Sun was difficult. Understanding the alignment post facto, during analysis, was similarly difficult, since the Sun was (by design) not visible behind the occulter, and there were no other celestial references visible in the FOV. We are developing a design adaptation that would allow us to track the position of the Sun relative to the occulter, to solve this problem in future iterations.

The final occulter assembly's stiffness was limited by the stiffness of the occulter rods.  We used 2~mm diameter
rods in CATEcor, and \new{in conjunction with the chosen 75cm length of the truss,} this choice imposed a roughly 10--15~Hz fundamental frequency for small perturbations; the frequency
was determined by the stiffness of the rods themselves rather than of the truss as a whole.  At this modest level of stiffness, wind 
shake was a significant issue even in light breezes.  Subsequent designs 
could use a combination of thicker rods and/or a mid-rod stiffener bracket to raise the stiffness, ideally into at least
the 20--30~Hz range.

While FDM 3D printing is a convenient process for prototyping, it is not required for implementing a coronagraph
of this general type.  Other fabrication methods, including (additive) resin printing and (subtractive) 
conventional machining, and other materials, including metals and stiffer plastics than the PETG used, provide much 
more precision and performance, and would improve both the diffractive/optical performance and 
stiffness of this demonstration design.

The offset between the occulter and aperture is not strictly limited to 75~cm as in the CATEcor design.
Commercial circular-cross-section carbon fiber rods afford excellent strength-to-mass ratio and stiffness, but
improved truss designs and scientific-grade materials afford yet greater strength.  Lengthening the 
aperture-occulter distance improves performance by reducing the inner diameter of the FOV, while also
reducing the Fresnel diffraction around the occulter.  The truss design imposes a necessary tradeoff between 
complexity of vignetting function and rigidity of the occulting structure; CATEcor is a first exploratory
cut at this novel design space, and is far from optimized.

\section{Conclusions}\label{sec:conclusions}

CATEcor is a proof-of-concept of a new type of instrument: a shaded-truss externally occulted coronagraph. We
designed it specifically to match the conditions of the 14-October-2023 annular eclipse, but the concept has 
applications beyond our initial deployment.  In particular, because
CATEcor was designed entirely with open-source CAD tools and implemented with materials and procedures 
available to 
amateur astronomers and advanced students, it demonstrates the feasibility of ``from scratch'' observations of the
solar corona for non-scientists, students, and amateur astronomers.  Further, CATEcor opens a new design space
of shaded-truss coronagraphs, with the potential to offer better performance and broader fields of view than
conventional designs.

\begin{acks}
This work was supported by the Citizen CATE 2024 project, and by Southwest Research Institute via internal funding.  Funding for Citizen CATE 2024 is provided by grants from the NSF (awards 2231658, 2308305, \& 2308306) and NASA (grants 80NSSC21K0798 \& 80NSSC23K0946).  The authors thank the
FreeCAD project and AutoDesk for making powerful and accessible CAD tools
available to the general public.
\end{acks}

\begin{authorcontribution}
DeForest analyzed, designed, and constructed the coronagraph. Seaton conceived
the idea of a low-cost externally occulted coronagraph for CATE.  Caspi leads the Citizen CATE 2024
project and provided design input, feedback, and direction.  Beasley provided
optical design input.  Kovac and Tosolini participated in integration and
deployment of the CATEcor coronagraph.  West, Patel, Davis, and Tosolini aided field-test of
the design.  All co-authors reviewed the work and
provided editing input. 
\end{authorcontribution}

\begin{fundinginformation}
CATEcor development and deployment was funded in part by the National Science Foundation under 
Grants 2231658 and 2308305, and in part by Southwest Research Institute under internal funding.  Funding for Citizen CATE 2024 was provided by grants from the NSF (awards 2231658, 2308305, \& 2308306) and NASA (grants  80NSSC21K0798 \& 80NSSC23K0946).
\end{fundinginformation}

\begin{materialsavailability}
The mechanical design is intended to be accessible to the public.  Printable .stl files are published on Thingiverse \citep{DeForest_2023}.
\end{materialsavailability}

\begin{ethics}
\begin{conflict}
The authors have no conflicts of interest.
\end{conflict}
\end{ethics}


\bibliographystyle{spr-mp-sola}
\bibliography{CATEcor-description}  

\end{document}